\documentclass[conference]{IEEEtran}
\usepackage{cite}
\usepackage{amsmath,amssymb,amsfonts}
\usepackage{algorithmic}
\usepackage{graphicx}
\usepackage{textcomp}
\usepackage{xcolor}
\usepackage{hyperref}
\usepackage{booktabs} 
\usepackage{balance}
\usepackage{mathptmx} 
\usepackage{placeins} 

\def\BibTeX{{\rm B\kern-.05em{\sc i\kern-.025em b}\kern-.08em
    T\kern-.1667em\lower.7ex\hbox{E}\kern-.125emX}}

\definecolor{darkblue}{rgb}{0.1,0.1,0.4}
\hypersetup{
  pdftitle={The Impact of User Location on Cookie Notices},
  pdfauthor={Rob van Eijk, Hadi Asghari, Philipp Winter, and Arvind Narayanan},
  pdfkeywords={Cookie Notice, Web Privacy Measurement, ePrivacy, GDPR, VPN},
  colorlinks=true,
  urlcolor=darkblue,
  linkcolor=darkblue,
  citecolor=darkblue
}

\begin{document}

\title{The Impact of User Location on Cookie Notices\\(Inside and Outside of the European Union)}

\author{\IEEEauthorblockN{Rob van Eijk}
\IEEEauthorblockA{\textit{Leiden University} \\
}
\and
\IEEEauthorblockN{Hadi Asghari}
\IEEEauthorblockA{\textit{TU Delft} \\
}
\and
\IEEEauthorblockN{Philipp Winter}
\IEEEauthorblockA{\textit{CAIDA, UC San Diego} \\
}
\and
\IEEEauthorblockN{Arvind Narayanan}
\IEEEauthorblockA{\textit{Princeton University} \\
}
}

\maketitle

\begin{abstract}
The web is global, but privacy laws differ by country. Which set of privacy rules do websites follow? We empirically study this question by detecting and analyzing cookie notices in an automated way. We crawl 1,500 European, American, and Canadian websites from each of 18 countries. We detect cookie notices on 40\% of websites in our sample. We treat the presence or absence of cookie notices, as well as visual differences, as  proxies for differences in privacy rules. Using a series of regression models, we find that the website's Top Level Domain explains a substantial portion of the variance in cookie notice metrics, but the user’s vantage point does not. This suggests that websites follow one set of privacy rules for all their users. There is one exception to this finding: cookie notices differ when accessing .com domains from inside versus outside of the EU. We highlight ways in which future research could build on our preliminary findings.

\end{abstract}

\begin{IEEEkeywords}
Cookie Notice, Web Privacy Measurement, ePrivacy, GDPR, VPN
\end{IEEEkeywords}

\section{Introduction}
In the European Union (EU), the General Data Protection Regulation (GDPR)~\cite{parliament_of_the_eu_and_the_council_regulation_2016} and the ePrivacy Directive (ePD)~\cite{parliament_of_the_eu_and_the_council_directive_2002} mandate websites to inform users, and ask for their consent, before placing or accessing tracking cookies on their machines (amongst other rules). In an attempt to harmonize these rules across EU countries, legislators and regulators have provided additional guidance on what these notices should specify and what constitutes real choice~\cite{article_29_working_party_guidelines_2018}. However, due to the process of translating the ePD into national legislation, differences remain among how member states enforce the privacy rules (see, e.g., Dumortier and Kosta~\cite{dumortier_eprivacy_2015}). This creates a potentially confusing situation for website operators.

Consider the following example: a \textit{Belgian} user visits the popular \textit{Dutch} news website \textit{nu.nl}. Nu.nl  has its servers hosted by Amazon \textit{Ireland} and is owned by the \textit{Finnish} headquartered Sanoma Group. Which consent and notice rules should nu.nl follow for this user? The Belgian, Dutch, Irish, or Finnish one? 
Under the GDPR, firms have some freedom in choosing their \textit{main establishment} (as stipulated by Article 4(16) and Recital 36). Therefore the question becomes an empirical one: which rule do firms follow in practice? Is it predetermined or does it change given the user's location?

There is anecdotal evidence that the location of the user plays a role in the tracking behavior of websites~\cite{davies_after_2019}. 
Previous studies have touched upon, and found, geographical differences in tracking for a limited number of countries~\cite{fruchter_variations_nodate,leenes_taming_2015}, but none have systematically studied this question.  Recent studies have found an increase in cookie notices after the GDPR took effect~\cite{libert_exposing_2015,libert_third-party_2018,libert_changes_2018}. However, they also did not specifically look at cookie notices per country.

We hypothesize that websites will simplify their decision process by following the rule of their main target audience, for instance by using their Top Level Domain (TLD) as a proxy for their audience. We actively test this hypothesis by fetching the country-specific top 100 websites from 18 different vantage points (15 EU countries plus Canada, Switzerland, and the United States) in January 2019---half a year after the GDPR took effect. We expect to find that differences in cookie notices are driven primarily by the TLD and not the vantage point. Our findings support this hypothesis (Section \ref{sec:findings}).

For our study we needed an automated way to capture cookie notices. Past studies have looked at some of the usability and legal aspects of notices, e.g., Borgesius et al.~\cite{borgesius_tracking_2017}, but to the best of our knowledge none have done so in an automated manner. For this purpose, we extended the OpenWPM framework~\cite{englehardt_web_2014} to automatically detect and record cookie notices on any given website, with the aid of a list of banner elements maintained by the ``I don't care about cookies'' browser extension~\cite{idontcare}. 

In sum, our paper contributes to the state of the art in three ways:

\begin{itemize}
    \item We develop and evaluate an automated technique to detect cookie notices;
    \item We investigate whether the vantage point influences cookie notices shown to users;
    \item We investigate whether TLDs are a good predictor for the cookie notice rules followed by a website.
\end{itemize}

Finally, we discuss the implications of our findings for future web privacy measurement research.
\section{Background}

\subsection{Cookie Notices}
Article 5(3) of the ePD requires prior informed consent for storage of, or for access to, information stored on a user's terminal equipment. Cookies stored on a user's browser fall under this article, so websites must inform users and ask if they agree before using cookies, with some exceptions: 

\textit{``Member States shall ensure that the storing of information, or the gaining of access to information already stored, in the terminal equipment of a subscriber or user is only allowed on condition that the subscriber or user concerned has given his or her consent, having been provided with clear and comprehensive information, in accordance with Directive 95/46/EC, inter alia, about the purposes of the processing. This shall not prevent any technical storage or access for the sole purpose of carrying out the transmission of a communication over an electronic communications network, or as strictly necessary in order for the provider of an information society service explicitly requested by the subscriber or user to provide the service.''}\footnote{Now that the GDPR is in force, the reference to the Directive 95/46/EC has been replaced by a reference to art. 6(1) of the GDPR.}
\cite{parliament_of_the_eu_and_the_council_directive_2002}

 The requirements for consent are clarified in Recital 32 of the GDPR. Furthermore, the Article 29 Working Party (Art. 29 WP) has given extensive guidance on consent under the GDPR~\cite{article_29_working_party_guidelines_2017} and the ePD~\cite{article_29_working_party_opinion_2012, article_29_working_party_working_2013,article_29_working_party_opinion_2014,article_29_working_party_opinion_2017}. For consent to be valid for data processing, it needs to meet five aspects: (1) given by clear affirmative act, (2) given freely, (3) be specific, (4) be informed, and (5) give an unambiguous indication of the user's agreement. 
 
 The European Data Protection Board (EDPB), which is the successor to the Art. 29 WP, has provided additional guidance on the interplay between the GDPR and the ePD~\cite{edpb_2019}. They offer the example of 
 ``a data broker engaged in profiling on the basis of information concerning the internet browsing behaviour of individuals, collected by the use of cookies, but which may also include personal data
obtained via other sources''. 
In such a case, ``the placing or reading of cookies \textit{must comply with the national provision transposing article 5(3) of the ePrivacy Directive}. Subsequent processing of personal data including personal data obtained by cookies \textit{must also have a legal basis under article 6 of the GDPR} in order to be lawful.'' (emphasis added)

 Cookie notices presented in the form of cookie banners and cookie-walls  are how these legal requirements are usually met in practice.

\subsection{Prior Work}
Several aspects of cookie notices have been studied so far. One aspect relates to the trackers, including the legal structure of ad-tech companies~\cite{libert_exposing_2015}; the prevalence~\cite{englehardt_online_2016} and prominence~\cite{binns_measuring_2018} of trackers; and their mergers and acquisitions~\cite{van_eijk_web_2018}. Trevisan et al.~\cite{trevisan_uncovering_2017} investigated 35,000 websites and reported that 65\% of websites did not respect the legal requirements for consent as set out by the ePD. 

Another aspect involves the end-user perspective---the human factor long recognized in security (e.g., by Riegelsberger et al.~\cite{riegelsberger_mechanics_2005}). Keith et al. explained the effects of privacy control complexity using feature fatigue theory \cite{keith_privacy_2014}; Schermer, Custers, and Van der Hof investigated the effectiveness of consent in data protection legislation \cite{schermer_crisis_2014}; And Leenes and Kosta \cite{leenes_taming_2015} focused on the question of giving users a more meaningful choice with regard to the collection and processing of their personal information. 

Recently, scholars have started to investigate the variation of cross-border web tracking with respect to the GDPR and differences in ePD implementations throughout EU member states. For instance, Fruchter et al. \cite{fruchter_variations_nodate} and Van Eijk \cite{van_eijk_web_2018} investigated variations in tracking in relation to the geographic location of visiting a website. Iordanou et al. \cite{iordanou_tracing_2018} investigated prevalence in cross-border tracking and found that a large fraction of tracking flows across borders are served by servers in neighboring EU countries, and that the majority of such tracking is well confined within EU member states.

\section{Methods}

\subsection{Measurement Platform}

We used the OpenWPM platform \cite{englehardt_online_2016} to collect our measurements. OpenWPM automates the crawling of websites using a web browser (simulating a real user), and stores tracking-related elements (such as cookies, redirects, and JavaScript calls) in an SQLite database. 

To simulate users from different countries, we used a Virtual Private Network (VPN) provider with vantage points in different countries. We also configured OpenWPM's browser locale to match the most popular language of the respective vantage point. Finally, we dockerized OpenWPM to simplify crawling through multiple VPN connections on one machine.

\subsection{Banner detection}
OpenWPM has no means of detecting cookie banners. We added functionality to OpenWPM to detect cookie banners by drawing on a crowd-sourced list of CSS elements.\footnote{Our code is available in the following GitHub branch: \url{https://github.com/hadiasghari/OpenWPM/blob/master/automation/Commands/utils/banner_utils.py}} This list is part of the browser extension ``I don't care about cookies''~\cite{idontcare} and contains over 9,000 CSS element names that are typically used for cookie banners, e.g., \texttt{\#cookieNotice}, \texttt{\#cookieScreen}, and \texttt{\#acceptCookies}. Once our code detects a potential cookie banner, it logs its dimensions (height and width), its location offset ($x$ and $y$ coordinates), and its inner HTML. Note that since our OpenWPM extension merely matches the names of CSS elements, it is possible that our data includes false positives, e.g., a website that happened to use the CSS element \texttt{\#cookieScreen} to display recipes for chocolate chip cookies. We evaluate this in Section \ref{sec:banner-eval}.

In addition to the presence or absence of a banner, we also analyze the height of the banner, the number of words, and the number of links/buttons. These are easily measured proxies for underlying substantive differences in notice and consent rules, such as the amount of information provided to the user and the ability to refuse consent. We defer a full analysis of substantive differences to future work. 

\subsection{Choice of Countries \& Websites}
Given that the GDPR and ePD are EU legislations, we chose websites operating primarily in fifteen EU countries\footnote{To keep the work manageable, from the 28 EU member states we chose those with more than eight million population: AT/Austria, BE/Belgium, CZ/Czech Republic, DE/Germany, ES/Spain, FR/France, GR/Greece, HU/Hungary, IT/Italy, NL/Netherlands, PL/Poland, PT/Portugal, RO/Romania, SE/Sweden, and UK/United Kingdom.}, as well as CA/Canada, US/United States, and CH/Switzerland as control.

We selected the websites from TLDs for each country in the study. (We used the country-code top level domain, or the ccTLD, for all countries except the US; for the US we used the .com TLD). As mentioned in the introduction, we hypothesized TLDs to be a decent proxy for the target audience of a website, and thus expected websites to implement the privacy rules for that country. We scoped the research to the top 100 websites active in each country, in part because following complex legislation requires resources that might not be available to smaller firms. Additionally, top websites would expect scrutiny in case of non-compliance. To determine  the top websites in each country (TLD), we used the \textit{Majestic} list \cite{majestic}. We found this list to represent the top EU websites much better than common alternatives (including Alexa) upon eyeballing the Dutch list\footnote{Turcios Rodriguez \cite{rodriguez2018} compares Majestic with several common site rankings (see Table 5), and recommends Majestic over the others, in part because it is used for search engine optimization by marketers. }.

\subsection{Crawl Details}
We conducted a pilot crawl in May 2018 to test our setup. We ran another crawl on January 7-9, 2019, approximately 7 months after the GDPR has come into effect, and the measurements reported here are based on this crawl. We use \textit{Mullvad} \cite{mullvad} as the VPN provider.\footnote{We originally used HideMyAss! provider but during this time issues were discovered with fake locations used by them \cite{khan18_vpn}.} 
We used the banner selector list from December 2018. Websites were crawled in a short time-span and in parallel from five Docker instances to reduce the temporal effects of ad-campaigns. 

The crawler failed on about 15\% of the websites in our sample. Upon inspection, the majority of the failures were due to two shortcomings of the Majestic list: the listing of generic second level domains which do not host a website (such as .ac.uk), and the listing of websites which return 400/500 errors even when accessed directly.\footnote{Such shortcomings are unfortunately common among the major top sites lists; \cite{letranco19} offers a recent analysis.} About 1\% of the failures are unspecified OpenWPM/Selenium crashes. After removing the failed measurements, we are left with 1,543 sites and 27,488 measurements in our dataset.

\section{Findings}
\label{sec:findings}
\subsection{Prevalence of Banners}
We detected a cookie banner on 40.2\% of the websites crawled (averaged over all vantage points). The median banner has a height of 86 pixels; a notice text of 31 words; and contains two links/buttons (e.g., linking to a cookie policy, an accept button, and sometimes a refuse button). 

\begin{figure}[t]
\centerline{\includegraphics[width=\linewidth]{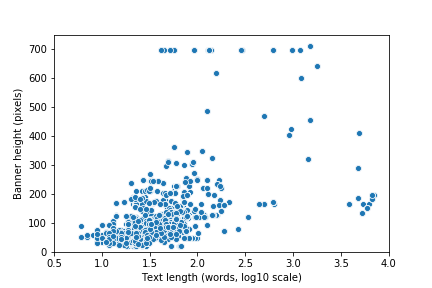}}
\caption{Scatter plot of cookie banner heights vs notice length for all websites crawled (5 outliers removed). Spearman rho=0.63 (p-value=0.00).}
\label{fig:scatterhl}
\end{figure}

Figure \ref{fig:scatterhl} shows the scatter-plot between banner height and text lengths.  Note that 700 pixels is the full screen height of the Firefox crawler, so a banner of 400 pixels fills up half the visible screen. The longer notice texts (e.g., \textit{kleinezeitung.at}) list all the third party trackers on the banner. (Such an implementation is unusual; most banners offer the full tracker list on other pages). 

Websites which show a cookie banner store a median of 4 third-party persistent cookies, while those without a banner store a median of 1 such cookie  This pattern is in line with the exemptions set by the ePD that functional non-tracking cookies do not need explicit consent.

\subsection{Banner Detector Evaluation}
\label{sec:banner-eval}
\textit{Detection Speed.} Matching the several thousand CSS selectors to detect a cookie banner took on average 18 seconds per site (st. dev. = 18 seconds); for the slowest site in our sample (\textit{youtube.com}) it took 117 seconds. When a banner is detected, the median number of HTML elements that match is one, with a maximum of seven (excluding two outliers). Multiple matches can either be (1) because of multiple banners (e.g., one at the top and one at the bottom of the page); or (2) because of nested elements. In both cases we take the height of the tallest element and the length of the longest text for the measure. 

\begin{table}[t]
\caption{Accuracy of the banner detector}
\label{tbl:accuracy}
\centering
\begin{tabular}{@{}l r r r@{}}
\toprule
\textbf{}       & \textbf{.ca sites} & \textbf{.es sites} & \textbf{.nl sites} \\ 
\midrule
True positive   & 11    & 40    &  48   \\
True negative   & 70    & 26    &  23   \\ 
False positive  & 0     & 1     &  0    \\
False negative  & 0     & 4     &  17   \\
Crawl failure (see section III-C)   & 19    & 29    &  12    \\
\bottomrule
\multicolumn{3}{l}{\textit{Crawls from the respective countries.  (VPN=CA/ES/NL)}} \\
\end{tabular}
\end{table}

\textit{Detection Accuracy.} We visually inspected the detected cookie notices for a subset of the crawls---Canadian websites (.ca TLD), Spanish websites (.es TLD), and Dutch websites (.nl TLD) as viewed from their respective countries (VPN=CA, ES, NL). Table \ref{tbl:accuracy} presents the results.  Based on this subset, we have a false positive rate (FPR) of less than 1\%, a false negative rate (FNR) of approximately 18\%, and a total accuracy of 91\%.  

The false negative rate is particularly high for the Netherlands.  The majority of the missed banners (14) were linked to websites using \textit{cookie-walls}---full screen overlays that need to be clicked before the user can view any content. Cookie-walls are not on the crowd-sourced CSS selector list (according to the creator of the extension, this is because they require a JavaScript action to hide). Missing cookie-walls is thus a limitation of our technique. 

The true positive regarding banners presented by Canadian websites and to Canadian users (e.g., \textit{ucalgary.ca}) were unexpected. They might be due to stricter notice rules in some Canadian provinces, or indicate an extra territorial impact of the GDPR (i.e., organizations wishing to simplify their privacy implementation applying the stricter rule).

\subsection{Effects of User Location (VPN)}
We now inspect how a user's location---the VPN country---effects cookie notices using visual inspection and using regression analysis. 


\begin{figure*}[htb]
\includegraphics[width=\linewidth]{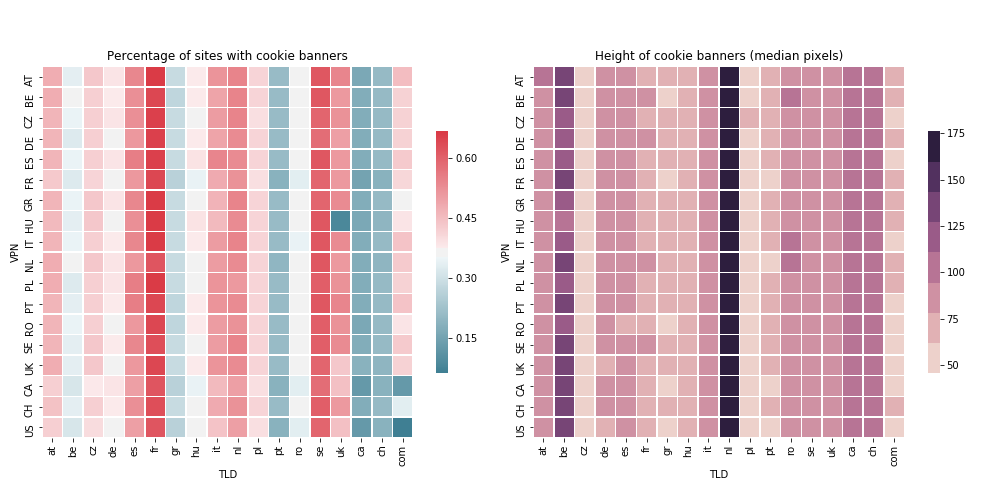}
\vspace*{-30pt}
\caption{Heatmaps showing banner prevalence and height across websites grouped by TLD and crawl vantage point. Row labels are ISO country codes.}
\label{fig:heatmap}
\end{figure*}


Fig. \ref{fig:heatmap}-left shows a heatmap for the prevalence of banners where we have grouped measurements by website TLD (columns) and the VPN used for that crawl (rows). The heatmap on the right shows median banner height using the same grouping. Patterns that appear in rows can be linked to VPN effects; patterns that appear in columns can be linked to TLDs.  The figure reveals strong TLD effects (i.e., clear differences among the columns);  it also reveals negligent VPN effects except for .com websites when viewed from outside of the EU. The vantage point effect for .com websites is in part because sites such as \textit{google.com} redirect the user to their local versions (e.g., \textit{google.nl}) based on geo-location.

Table \ref{tbl:regbanner} investigate the  relationship between the existance of a cookie and TLDs and VPNs using Logit regression models. The top box compares the models with different (fixed-effect) predictors using the Akaike information criterion (AIC). The model with the lowest AIC score fits the data better; the AIC includes a term that penalizes models having many parameters, which discourages overfitting. As seen in the table, the model with the lowest AIC is the one with predictors for all TLDs, plus a predictor for whether the VPN was from inside or outside of the EU, and an interaction term between VPN\_EU and .com TLD. (The interaction term means that the VPN location primarily impacts .com websites). 

The bottom box lists the coefficient values as well as the equivalent increase or decrease of \textit{odds ratio}. A .at website, for example, has an increased 47\% odds of having a cookie banner  compared to the baseline. (We chose .be websites as the baseline since they were in the middle of the pack on prevalence). A .com website has a decreased 70\% odds of having a banner compared to the baseline if visited from outside of the EU, and 32\% increased odds if visited from inside the EU.

\begin{table}[htb]
\vspace*{-5pt}
\caption{\label{tbl:regbanner}Regression Analysis}
\centering
\begin{tabular}{@{}l r r@{}}
\toprule
\textbf{Model}                & \textbf{AIC}    \\
\midrule
banner\_exists \~ {} VPNs        &    34,242              \\
banner\_exists \~{}  TLDs      & 31,874               \\
banner\_exists \~{}  TLDs + VPNs      & 31,890               \\
banner\_exists \~{}  TLDs + VPN\_EU    & 31,868               \\
banner\_exists \~{}  TLDs + VPN\_EU + VPN\_EUxTLD\_com    & \textbf{31,855}               \\
\bottomrule
\end{tabular}
\begin{tabular}{@{}l r r r@{}}
\cr
\multicolumn{4}{c}{\textsc{Coefficient for Best Model}} \\ 
\toprule
\textbf{Predictor}                & \textbf{Coefficient} & \textbf{(95\% conf. int.)}    & \textbf{Odds Ratio} \\
\midrule
TLD[.at] 	&	 0.39 &  $\pm 0.15$  	&	 $+47$\% 	\\
TLD[.ca] 	&	 $-1.66$ & $\pm 0.21$   	&	 $-81$\% 	\\
TLD[.ch] 	&	 $-0.83$ & $\pm 0.16$  	&	 $-56$\% 	\\
TLD[.cz] 	&	 0.20 & $\pm 0.14$ 	&	 $+23$\% 	\\
TLD[.de] 	&	 $-0.69$ & $\pm 0.16$  	&	 $-50$\% 	\\
TLD[.es] 	&	 0.27 & $\pm 0.15$  	&	 $+31$\% 	\\
TLD[.fr] 	&	 1.42 & $\pm 0.15$  	&	 $+313$\% 	\\
TLD[.gr] 	&	 $-0.26$ & $\pm 0.15$   	&	 $-23$\% 	\\
TLD[.hu] 	&	 $-0.23$ & $\pm 0.15$  	&	 $-20$\% 	\\
TLD[.it] 	&	 0.62 & $\pm 0.15$  	&	 $+85$\% 	\\
TLD[.nl] 	&	 0.48 & $\pm 0.14$  	&	 $+62$\% 	\\
TLD[.pl] 	&	 0.27 & $\pm 0.16$  	&	 $+31$\% 	\\
TLD[.pt] 	&	 $-0.41$ & $\pm 0.16$  	&	 $-34$\% 	\\
TLD[.ro] 	&	 $-0.39$ & $\pm 0.17$  	&	 $-32$\% 	\\
TLD[.se] 	&	 0.36 & $\pm 0.15$  	&	 $+43$\% 	\\
TLD[.uk] 	&	 0.85 & $\pm 0.15$  	&	 $+134$\% 	\\
TLD[.com] 	&	 $-1.21$ & $\pm 0.40$  	&	 $-70$\% 	\\
VPN\_EU 	&	 0.07 & $\pm 0.07$  	&	 $+7$\% 	\\
VPN\_EUxTLD\_com 	&	 0.75  & $\pm 0.40$  	&	 $+112$\% 	\\
Intercept 	&	 $-0.85$ & $\pm 0.12$  	&	  	\\
\bottomrule
\multicolumn{4}{l}{\textit{MLE Logit regression, N=27150, df=19. Baseline TLD=.be}} \\ 
\multicolumn{4}{l}{\textit{Pseudo $R^2$ (McFadden)=0.07, LLR p-value=0.00}} \\
\end{tabular}
\end{table}

In short, the regression results reveal the same as the  heatmap. Websites appear to follow notices requirements based on their expected audience (=ccTLD), and not individual user location (=VPN), with the exception of .com websites. 


\subsection{Differences Across TLDs and Time}
The results of the previous subsection also reveal substantial differences in banner statistics across TLDs. Since we concluded that for ccTLDs the vantage point can be ignored, we aggregate the measurements for each TLD, and present the results in Table \ref{tbl:tldstats}. 

What particularly stands out is the taller height of banners and the existence of one additional link (like a refuse button) for  Belgian and Dutch TLDs. These and other differences among TLDs likely reflect differences in the regulatory approach of the national data protection authorities (DPA), such as additional guidance or enforcement capacity. 

Regarding temporal differences, we compared statistics from our pilot and main crawls (the pilot was done immediately after the GDPR applicability and the main crawl seven months after). 
We found an increase from 32\% to 40\% in the number of cookie banners detected. This increase may be due to an increase in cookie banner implementations due to the GDPR; it may alternatively be due to improvements in the crowd-sourced selector list. The median number of banner links also increased from one to two, a change in sync with the stricter consent requirements. 

These findings are speculative given our limited experimental setup and analysis. They are, however, interesting points for further research.  

\begin{table}[!b]
\caption{\label{tbl:tldstats}Cookie banner statistics per TLD (median)}
\centering
\begin{tabular}{@{}l r r r r l@{}}
\toprule
\textbf{TLD $\dag$}    & \textbf{Freq. (\%)} & \textbf{Height (px)} & \textbf{Words} & \textbf{Links $\dag\dag$} & \textbf {Language $\ddag$} \\
\midrule
.at & 47 & 95 & 33  & 2 & German \\
.be & 34 & 129 & 32 & \textbf{3} & Flemish \\
.cz & 44 & 48 & 21 & 2 & Czech \\
.de & 39 & 80 & 27 & 2 & German \\
.es & 56 & 84 & 37 & 2 & Spanish \\
.fr & 67 & 77 & 45 & 2 & French \\
.gr & 29 & 63 & 26 & 2 & Greek \\
.hu & 39 & 72 & 21 & 2 & Hungarian \\
.it & 53 & 93 & 38 & 2 & Italian \\
.nl & 54 & 174 & 101 &  \textbf{3} & Dutch \\
.pl & 41 & 62 & 29 & 2 & Polish \\
.pt & 21 & 64 & 42 & 2 & English? \\
.ro & 36 & 94 & 28 & 2 & Romanian \\
.se & 62 & 92 & 27 & 2 & Swedish \\
.uk & 52 & 86 & 32 & 2 & English \\
.ca & 18 & 96 & 43 & 2 & English \\
.ch & 21 & 100 & 32 & 2 &  English? \\
\bottomrule
\multicolumn{6}{l}{\textit{$\dag$ Excluding .com since there the vantage point matters}} \\
\multicolumn{6}{l}{\textit{$\dag\dag$ Includes both links \& buttons}} \\ 
\multicolumn{6}{l}{\textit{$\ddag$ The main language is detected using the TextBlob library}} \\ 
\end{tabular}
\end{table}

\section{Discussion}
In this short paper we looked at how the multiplicity of EU data privacy laws might be interpreted by websites when accessed from different countries. We found support for our hypothesis that in the context of notice and consent rules, companies simplify their implementation efforts by using the website ccTLD, and not the user location, to decide which rules to follow. The exception are .com websites, where the presence of a cookie banner depends on whether they are accessed from inside or outside of the EU---the odds of seeing a cookie banner increases by 102\% when using a vantage point in the EU.

Some of the differences we observed  may be the result of variations in regulatory approach---e.g., taller banners on Belgian and Dutch websites, cookie notices on Canadian websites for Canadian users, and an increase in the number of links on cookie banners since the applicability of the GDPR. The findings on Dutch banners links to the work of Leenes and Kosta~\cite{leenes_taming_2015} who in 2015 described Dutch cookie law as a ``failure'' because it didn't offer users real consent options---a point which has since led to stricter notice guidelines by the Dutch DPA.\footnote{For an example of these guidelines, see the public statement by the Dutch DPA (in Dutch):  \url{https://autoriteitpersoonsgegevens.nl/sites/default/files/atoms/files/normuitleg_ap_cookie-walls.pdf}. In short, websites that give visitors conditional access to their site with a ``take-it-or-leave-it'' cookie-wall do not comply with the GDPR. Prior consent with a cookie-wall is ``not valid because website visitors are refused access to the website without consent. Under the GDPR, consent is not `free' if someone has no real or free choice, or if the person cannot refuse to give consent without adverse effects.''}

We view our paper as a starting point. Future research can more deeply explore regulatory differences within EU countries and their relationship to websites' notice and consent implementations, both from a legal perspective and an empirical perspective. Empirical directions that we did not explore include analyzing the text of websites' privacy policies and interacting with banners---by following their links or measuring cookie placement. 

We hope that our technique for automatically detect cookie notices will be useful to other researchers. We found it to be  effective (FPR under 1\%, FNR approximately 20\%) with the main limitation being the detection of cookie-walls. 

Finally, our work has implications for the methodology of web privacy measurement, a line of research that involves analyzing websites to study privacy-impacting behaviors, such as third-party cookies or browser fingerprinting. An open question is the number of vantage points or countries that are needed for comprehensive measurements, given that websites might localize their content, ads, and trackers based on the location of the user. 

Our finding of a vantage point difference for .com crawls, along with anecdotal evidence from news websites treating EU and non-EU users differently, leads us to recommend that WPM projects \emph{use at least two vantage points for their crawls}---one within the EU and one outside---to increase measurement accuracy. While our measurement set was limited to the top websites and the major EU countries, we expect the pattern to hold for websites further down the tail as well, since the desire to simplify implementation rules will be shared by smaller organizations.

\section*{Acknowledgements}
The authors thank Steven Englehardt, Elsa Rebeca Turcios Rodríguez, Martijn Warnier, and the anonymous reviewers for their valuable input and feedback during this research project.

\balance
\bibliographystyle{ieeetr}
\bibliography{CookieNotices.bib}

\begin{thebibliography}{10}

\bibitem{parliament_of_the_eu_and_the_council_regulation_2016}
{Parliament of the EU and the Council}, ``Regulation ({EU}) 2016/679 of the
  {European} {Parliament} and of the {Council} of 27 {April} 2016 on the
  protection of natural persons with regard to the processing of personal data
  and on the free movement of such data, and repealing {Directive} 95/46/{EC}
  ({General} {Data} {Protection} {Regulation}),'' 2016.

\bibitem{parliament_of_the_eu_and_the_council_directive_2002}
{Parliament of the EU and the Council}, ``Directive 2002/58/{EC} of the
  {European} {Parliament} and of the {Council} of 12 {July} 2002 concerning the
  processing of personal data and the protection of privacy in the electronic
  communications sector ({Directive} on privacy and electronic
  communications),'' July 2002.

\bibitem{article_29_working_party_guidelines_2018}
{Article 29 Working Party}, ``Guidelines on consent under {Regulation}
  2016/679,'' 2018.
\newblock Published: WP 259 rev.01.

\bibitem{dumortier_eprivacy_2015}
J.~Dumortier and E.~Kosta, ``{ePrivacy} {Directive}: assessment of
  transposition, effectiveness and compatibility with proposed {Data}
  {Protection} {Regulation},'' 2015.
\newblock Published: SMART 2013/007116.

\bibitem{davies_after_2019}
J.~Davies, ``After {GDPR}, {The} {New} {York} {Times} cut off ad exchanges in
  {Europe} — and kept growing ad revenue..''
  https://digiday.com/media/new-york-times-gdpr-cut-off-ad-exchanges-europe-ad-revenue/,
  2019.
\newblock Published: Blogpost, Digiday UK, Accessed: 2019-03-24.

\bibitem{fruchter_variations_nodate}
N.~Fruchter, H.~Miao, S.~Stevenson, and R.~Balebako, ``Variations in tracking
  in relation to geographic location,'' in {\em Proceedings of W2SP 2015},
  IEEE, 2015.

\bibitem{leenes_taming_2015}
R.~Leenes and E.~Kosta, ``Taming the cookie monster with dutch law–a tale of
  regulatory failure,'' {\em Computer Law \& Security Review}, vol.~31, no.~3,
  pp.~317--335, 2015.

\bibitem{libert_exposing_2015}
T.~Libert, ``Exposing the hidden web: {An} analysis of third-party http
  requests on 1 million websites,'' {\em International Journal of
  Communication}, vol.~9, pp.~3544--3561, 2015.

\bibitem{libert_third-party_2018}
T.~Libert and L.~Graves, ``Third-party web content on {EU} news sites.
  {Potential} challenges and paths to privacy improvement.''
  \url{https://timlibert.me/pdf/Libert_Nielsen-2018-Third_Party_Content_EU_News_GDPR.pdf},
  2018.
\newblock Published: Factsheet. Accessed: 2019-03-24.

\bibitem{libert_changes_2018}
T.~Libert, L.~Graves, and R.~K. Nielsen, ``Changes in third-party content on
  {European} news websites after {GDPR}.''
  \url{https://timlibert.me/pdf/Libert_et_al-2018-Changes_in_Third-Party_Content_on_EU_News_After_GDPR.pdf},
  2018.
\newblock Published: Factsheet. Accessed: 2019-03-24.

\bibitem{borgesius_tracking_2017}
F.~J.~Z. Borgesius, S.~Kruikemeier, S.~C. Boerman, and N.~Helberger, ``Tracking
  {Walls}, {Take}-{It}-{Or}-{Leave}-{It} {Choices}, the {GDPR}, and the
  {ePrivacy} {Regulation},'' {\em European Data Protection Law Review}, vol.~3,
  no.~3, pp.~353--368, 2017.

\bibitem{englehardt_web_2014}
S.~Englehardt, C.~Eubank, P.~Zimmerman, D.~Reisman, and A.~Narayanan, {\em Web
  {Privacy} {Measurement}: {Scientific} principles, engineering platform, and
  new results}.
\newblock Princeton, NJ: Center for Information Technology Policy, 2014.
\newblock Published: Draft, Jun 1, 2014.

\bibitem{idontcare}
D.~Kladnik, ``I don't care about cookies.''
  \url{https://www.i-dont-care-about-cookies.eu/}, 2018.
\newblock Published: Browser Plugin version2.9.8. Accessed: 2019-01-22.

\bibitem{article_29_working_party_guidelines_2017}
{Article 29 Working Party}, ``Guidelines on consent under {Regulation}
  2016/679,'' 2017.
\newblock Published: WP 259 rev.01.

\bibitem{article_29_working_party_opinion_2012}
{Article 29 Working Party}, ``Opinion 04/2012 on {Cookie} {Consent}
  {Exemption},'' 2012.
\newblock Published: WP 194.

\bibitem{article_29_working_party_working_2013}
{Article 29 Working Party}, ``Working {Document} 02/2013 providing guidance on
  obtaining consent for cookies,'' 2013.
\newblock Published: WP 208.

\bibitem{article_29_working_party_opinion_2014}
{Article 29 Working Party}, ``Opinion 09/2014 on the application of {Directive}
  2002/58/{EC} to device fingerprinting,'' 2014.
\newblock Published: WP 224.

\bibitem{article_29_working_party_opinion_2017}
{Article 29 Working Party}, ``Opinion 01/2017 on the {Proposed} {Regulation}
  for the {ePrivacy} {Regulation} (2002/58/{EC}),'' 2017.
\newblock Published: WP 247.

\bibitem{edpb_2019}
{European Data Protection Board}, ``{Opinion 5/2019 on the interplay between
  the ePrivacy Directive and the GDPR, in particular regarding the competence,
  tasks and powers of data protection authorities},'' 2019.
\newblock Adopted on 12 March 2019.

\bibitem{englehardt_online_2016}
S.~Englehardt and A.~Narayanan, ``Online {Tracking}: {A} 1-million-site
  {Measurement} and {Analysis},'' pp.~1388--1401, ACM Press, 2016.

\bibitem{binns_measuring_2018}
R.~Binns, J.~Zhao, M.~Van~Kleek, and N.~Shadbolt, ``Measuring third party
  tracker power across web and mobile,'' {\em arXiv:1802.02507 [cs]}, Feb.
  2018.
\newblock 00004 arXiv: 1802.02507.

\bibitem{van_eijk_web_2018}
R.~J. Van~Eijk, {\em Web {Privacy} {Measurement} in {Real}-{Time} {Bidding}
  {Systems}. {A} {Graph}-{Based} {Approach} to {RTB} system classification.}
\newblock {PhD} {Thesis}, eLaw - Center for Law and Digital Technologies of the
  Institute for the Interdisciplinary Study of the Law, Faculty of Law of
  Leiden University \& the Dual PhD Centre, Faculty of Governance and Global
  Affairs of Leiden University, 2018.

\bibitem{trevisan_uncovering_2017}
M.~Trevisan, S.~Traverso, H.~Metwalley, and M.~Mellia, {\em Uncovering the
  {Flop} of the {EU} {Cookie} {Law}}.
\newblock arXiv preprint arXiv:1705.08884, 2017.

\bibitem{riegelsberger_mechanics_2005}
J.~Riegelsberger, M.~A. Sasse, and J.~D. McCarthy, ``The mechanics of trust:
  {A} framework for research and design,'' {\em International Journal of
  Human-Computer Studies}, vol.~62, no.~3, pp.~381--422, 2005.

\bibitem{keith_privacy_2014}
M.~Keith, C.~Maynes, P.~Lowry, and J.~Babb, ``Privacy fatigue: {The} effect of
  privacy control complexity on consumer electronic information disclosure,''
  2014.

\bibitem{schermer_crisis_2014}
B.~W. Schermer, B.~Custers, and S.~van~der Hof, ``The crisis of consent: {How}
  stronger legal protection may lead to weaker consent in data protection,''
  {\em Ethics and Information Technology}, vol.~16, no.~2, pp.~171--182, 2014.

\bibitem{iordanou_tracing_2018}
C.~Iordanou, G.~Smaragdakis, I.~Poese, and N.~Laoutaris, ``Tracing cross border
  web tracking,'' in {\em Proceedings of the Internet Measurement Conference
  2018}, pp.~329--342, ACM, 2018.

\bibitem{majestic}
A.~Chudnovsky and S.~Pitchford, ``Majestic: Marketing search engine and {SEO}
  backlink checker.'' \url{https://majestic.com/}.
\newblock Accessed: 2019-01-22.

\bibitem{rodriguez2018}
E.~Turcios~Rodriguez, ``Tracking cookies in the european union, an empirical
  analysis of the current situation,'' Master's thesis, Faculty of Technology,
  Policy and Management, Delft University, the Netherlands, 2018.

\bibitem{mullvad}
F.~Str\"{o}mberg and D.~Berntsson, ``Mullvad {VPN}.''
  \url{https://mullvad.net}.
\newblock Accessed: 2019-01-22.

\bibitem{khan18_vpn}
M.~T. Khan, J.~DeBlasio, G.~M. Voelker, A.~C. Snoeren, C.~Kanich, and
  N.~Vallina-Rodriguez, ``An empirical analysis of the commercial vpn
  ecosystem,'' in {\em Proceedings of the Internet Measurement Conference
  2018}, IMC '18, (New York, NY, USA), pp.~443--456, ACM, 2018.

\bibitem{letranco19}
V.~Le~Pochat, T.~Van~Goethem, W.~Joosen, V.~Le~Pochat, T.~Van~Goethem,
  W.~Joosen, G.~Franken, T.~Van~Goethem, W.~Joosen, V.~Rimmer, {\em et~al.},
  ``Tranco: A research-oriented top sites ranking hardened against
  manipulation,'' in {\em Proceedings of the 26th Annual Network and
  Distributed System Security Symposium (NDSS 2019)}, 2019.

\end{thebibliography}

\end{document}